\documentclass[aps,prd,twocolumn,nofootinbib]{revtex4-1}
\usepackage{epsfig}
\usepackage[colorlinks,linkcolor=blue,anchorcolor=blue,citecolor=blue,urlcolor=blue,breaklinks=true]{hyperref}
\usepackage{graphics}
\usepackage{slashed}
\usepackage{color}
\usepackage{amsfonts}
\usepackage{amsmath}
\usepackage{extarrows}
\usepackage{amssymb}
\begin{document}

\title{Evaluation of pion-nucleon sigma term in Dyson-Schwinger equation approach of QCD}

\author{Jing-Hui Huang$^{1}$}\email{Email:jinghuihuang@cug.edu.cn}
\author{Ting-Ting Sun$^{1}$} \email{Email:tingtingsun@cug.edu.cn}
\author{Huan Chen$^{1}$}\email{Email:huanchen@cug.edu.cn}
%
%\author{Yu-xin Liu $^{2}$}\email{Email:yxliu@pku.edu.cn}

\address{$^{1}$ School of Mathematics and Physics, China University of Geosiciences, Lumo Road 388, 430074 Wuhan, China.}
%
%\address{$^{2}$ Department of Physics and State Key Laboratory of Nuclear Physics and Technology, Peking University, Beijing 100871,China.}

\begin{abstract}
We calculate the variation of the chiral condensate in medium with respect to the quark chemical potential and evaluate pion-nucleon sigma term via the Hellmann-Feynman theorem. The variation of chiral condensate in medium is obtained by solving the truncated Dyson-Schwinger equation for quark propagator at finite chemical potential, with various ans\"{a}tz for the quark-gluon vertex and gluon propagator.  We obtain the value of the sigma term $\sigma_{\pi N}$= 62(1) MeV, where the (1) represents the systematic error due to our different ans\"{a}tz for the quark-gluon vertex and gluon propagator.
Our result favors a relatively large value and is rather consistent with the recent data obtained by analyzing  pion-nucleon scattering and pionic-atom experiments.

%\bigskip

%\noindent Key-words:  Dyson-Schwinger equations,pion-nucleon sigma term , the Hellmann-Feynman theorem

%\bigskip

%\noindent PACS Number(s): 14.20.Dh, 14.40.Aq, 11.10.Lm, 12.39.Fe

\end{abstract}
\maketitle

%\vbox{}

\section{Introduction}
\label{intro}

pion-nucleon sigma term $\sigma_{\pi N}^{}$ is of fundamental importance for understanding the chiral symmetry breaking effects in
nucleon~\cite{Bednar2018,Dmitrasinovic2016} and the origin of the mass of observable matter ~\cite{Roberts-Hadron-2,Leutwyler2015}. Recently, special attention has been paid to $\sigma_{\pi N}^{}$,  since it is also significant in searches for the Higgs boson, supersymmetric particles, cold dark matter\cite{Bishara2017,Bottino2008,Giedt2009}, and CP violation \cite{Yamanaka:2017mef,Yanase:2018qqq}.
$\sigma_{\pi N}$ can be obtained indirectly in experiments, such as pion-nucleon scattering or pionic-atom experiments\cite{Pavan1999,Hoferichter2015,Ruiz2018,Friedman:2019zhc}.  Several recent analyses ~\cite{Hoferichter2015,Ruiz2018,Friedman:2019zhc} gave $50$~MeV~$<\sigma_{\pi N}<70$~MeV, which is relatively larger than the widely used value $ \sigma_{\pi N}\simeq 45$~MeV~\cite{Gasser1991}.
In particular, Refs.~\cite{Hoferichter2015,Ruiz2018} gave a value around $60$~MeV with quite small error bars.
Theoretically pion-nucleon sigma term could be calculated in chiral perturbation theory~\cite{Camalich2012,Fernando2018,Ren2018,RuizdeElvira2017},
lattice QCD~\cite{Ohki2008,Bail2016,Abdel2016,Yang2016,Varnhorst2018,Yamanaka2018},
Dyson-Schwinger Equation(DSE) approach of QCD~\cite{Chang2005,Flambaum2005,Yamanaka:2014lva} and various other models~\cite{Olsson2000,Hite2005,Huang2007,Bratkovskaya2017}. However, theoretical results vary widely with different methods. Notably, the values from lattice QCD are around 30 to 40 MeV, which are much smaller than the above experimental analyses. Conversely, some other works gave relatively large values, even up to 80 MeV ~\cite{Hite2005} or 95 MeV~\cite{Yamanaka:2014lva}.
% The article~\cite{Yamanaka:2014lva} in the Dyson-Schwinger formalism which calculating the gluon dressing effect to the quark scalar density gives the $ \sigma_{\pi N}\simeq 95 MeV$.
Thus, further effort is needed in the theoretical calculations of the sigma term. In this work, we evaluate pion-nucleon sigma term in the DSE approach of QCD, via the Hellmann-Feynman theorem.

Theoretically, pion-nucleon sigma term $\sigma_{\pi N}^{}$ is usually written via the Hellmann-Feynman theorem as

\begin{equation}
\label{eqn:sigmaterm-definition}
\sigma_{\pi N} = m_{q} \frac{\partial M_{N}^{}}{\partial m_{q}} \, ,
\end{equation}
where $M_{N}^{}$ is the nucleon mass and $m_{q}$ is the average current-quark-mass for u and d quarks (see Sec.~\ref{SecIII} for details).

It has been known that the nucleon mass $M_{N}^{}$ comes almost entirely from the dynamical chiral symmetry breaking (DCSB)
(see, e.g., Ref.~\cite{Roberts-Hadron-2}).
It has also been known that the DSEs of QCD provide a natural approach to investigate the DCSB and the chiral symmetry restoration in vacuum  (see, e.g., Refs.~\cite{Roberts1990,Chang2007,Wang2013,QCDPT-DSE21-2}), in hot medium (see, e.g., Refs.~\cite{DSE-1-2,QCDPT-DSE11,QCDPT-DSE12-4,QCDPT-DSE12-5,QCDPT-DSE13,QCDPT-DSE23}), in cold dense matter (see, e.g., Refs.~\cite{YuanChen2006,Chen2008,Chen2009,Chen:2011my,Chen:2016ran,QCDPT-DSE12-7}),
as well as the properties of hadrons (see, e.g., Refs.~\cite{Maris2003,Chang2009-1,CLR,Roberts-Hadron-3,Eichmann2016,GCL2017,Eichmann:2016yit}). %

Inspired by the above-mentioned successes of the DSE approach, we restudy the pion-nucleon sigma term in the DSE approach with both the widely used rainbow approximation and the Ball-Chiu vertex~\cite{BC19811,BC1981} for the effective quark-gluon vertex, and two different infrared-dominant models for the effective interaction.

The paper is organized as follows. In Sec.~\ref{SecII}, the truncation scheme of the DSE for the quark propagator in vacuum and at finite chemical potential is given. In Sec.~\ref{SecIII}, we briefly describe the  method for evaluating
pion-nucleon sigma term $\sigma_{\pi N}^{}$ via the DCSB in medium (more explicitly, the chiral condensate in medium).
Then, the numerical results are given in Sec.~\ref{SecIV}. Finally, we summarize our work and conclude with a brief remark in Sec.~\ref{SecV}.

\section{Dyson-Schwinger Equation for the quark propagator }
\label{SecII}
Our calculation is based on the quark propagator at finite chemical potential $S(p ; \mu)$, 
which satisfies the Dyson-Schwinger equation
\begin{eqnarray}
S(p ; \mu)^{-1}&= &Z_2 (i\gamma\cdot \tilde{p}+m_{q}) % \nonumber \\
 +  Z_1 g^2(\mu)\int \frac{d^4q}{(2\pi)^4}
 \nonumber \\ & &\times\!D_{\rho\sigma}(k;\mu) \frac{\lambda^a}{2}\gamma_\rho S(q;\mu)
\Gamma^a_\sigma(q,p;\mu) \, ,
\label{gendse}
\end{eqnarray}
where $\tilde p=(\vec p,p_4+i\mu)$, $k=p-q$, $D_{\rho\sigma}(k ;\mu)$ is the full gluon propagator, $\Gamma^a_\sigma(q,p;\mu)$ is the dressed quark-gluon vertex, $Z_{1}$ is the renormalization constant for the quark-gluon vertex, and $Z_{2} $ is the quark wave-function normalization constant. The general structure of the quark propagator at finite chemical potential can be written as
\begin{eqnarray}
S^{-1}(p;\mu)&=&i\vec{\gamma}\cdot
\vec{p}A(\vec{p}^2,p_4;\mu)+i\gamma_4\tilde{p_4}C(\vec{p}^2,p_4;\mu)
\nonumber \\ & & + B(\vec{p}^2,p_4;\mu) \, ,
\end{eqnarray}
where $A(\vec{p}^2,p_4;\mu)$, $B(\vec{p}^2,p_4;\mu)$,$C(\vec{p}^2,p_4;\mu)$ are scalar functions of $p^2$ and $p_4$, while in vacuum
$A(\vec{p}^2,p_4;\mu=0)=C(\vec{p}^2,p_4;\mu=0)=A_0(p^2)$, $B(\vec{p}^2,p_4;\mu)=B_0(p^2)$.
\begin{equation}
S^{-1}(p) = i\gamma\cdot pA(p^2) + B(p^2) \, .
\end{equation}

We solve Eq.~(\ref{gendse}) with models of the gluon propagator and the quark-gluon vertex, which describe meson properties in vacuum well in the symmetry-preserving Dyson-Schwinger equation and Bethe-Salpeter equation (BSE) schemes (see, e.g., Refs.~\cite{Chang2009,BSE,Fischer2009,Binosi:2014aea}).
In vacuum they are usually taken as
\begin{equation}
 { Z_{1}g^2 D_{\rho \sigma}(k) \Gamma_\sigma^a(q,p)}  =
{\cal G}(k^2)D^0_{\rho
\sigma}(k)\frac{\lambda^a}{2}\Gamma_{\sigma}(p,q) \, ,
\label{KernelAnsatz}
\end{equation}
where $D^{0}_{\rho \sigma}(k)=\frac{1}{k^2}\Big[\delta_{\rho\sigma}-\frac{k_\rho k_\sigma}{k^2} \Big]$ is the Landau-gauge free gauge-boson propagator, ${\cal G}(k^2)$ is a model effective interaction, and $\Gamma_{\sigma}(q,p)$ is the effective quark-gluon vertex. 
Since the chemical potential only appears explicitly in the DSE of quark propagator, its effects on the gluon propagator and quark-gluon vertex are indirect. We expect these effects at lower order, except in the Ball-Chiu (BC) vertex \cite{BC19811,BC1981} ans\"{a}tz, where we introduce the quark-gluon vertex modification via the nonperturbative Ward-Takahashi identity.
The extended form of the BC vertex at finite chemical potential was given in Ref.~\cite{Chen2008}
\begin{eqnarray}
i\Gamma_\sigma^{BC}(q,p;\mu) = && i \Sigma_A(q,p;\mu) \gamma^\perp_\sigma+ i \Sigma_C(q,p;\mu) \gamma^\|_\sigma \nonumber \\
&&+ (\tilde q+\tilde p)_\sigma[\frac{i}{2}\gamma^\perp\cdot (\tilde q+\tilde p) \Delta_A( q, p;\mu)] \nonumber \\
&&+ (\tilde q+\tilde p)_\sigma [ \frac{i}{2}\gamma^\|\cdot (\tilde q+\tilde p) \Delta_C(q, p;\mu) \nonumber \\
 &&+(\tilde q+\tilde p)_\sigma  \Delta_B(q, p;\mu) ] \, , \label{bcvtxmu}
\end{eqnarray}
where $\gamma^\| = (\vec 0,\gamma_4)$, $\gamma^\perp = \gamma - \gamma^\| $, $F=A$, $B$, $C$.
\begin{eqnarray}
 \label{bcvtxmu2}
 \Sigma_F(q,p;\mu) = && \frac{1}{2} \left[ F(\vec q^2,q_4;\mu)+F(\vec p^2,p_4;\mu)\right] \, , \nonumber\\
 \Delta_F(q, p;\mu) =&& \frac{F(\vec q^2,q_4;\mu)-F(\vec p^2,p_4;\mu)}{\tilde q^2-\tilde p^2} \, . \nonumber\label{bcvtxmu2}
\end{eqnarray}
As a comparison, we also investigate the rainbow approximation for the vertex:
\begin{equation}
\label{vertex} \Gamma^{RB}_\sigma(q,p;\mu) = \gamma_{\sigma}^{} \, .
\end{equation}

For the model effective interaction, we employ two infrared-dominant models, denoted as the "GS" and the `"QC" models,
which only express the long-range behavior of the renormalization-group-improved Maris-Tanday model~\cite{Maris1999},
and the Qin-Chang(QC) model~\cite{Alkofer:2002bp,QC2011}. 
Though the ultraviolet parts of the models ensure the correct perturbative behavior,
they are not essential in describing nonperturbative physics, e.g. spectrum of light mesons ~\cite{Alkofer:2002bp,Eichmann2016}.
Herein we neglect them since our work is based on the competition between the chiral condensate and quark number density in the infrared region ~\cite{Chen2009}.
The two models are expressed as:
\begin{equation}
\label{IRGs} 
{\cal G}^{GS}(k^{2})= \frac{4\pi^2}{\omega^6} \, D\,k^{2}\, {\rm e}^{-k^{2}/\omega^2} \, ,
\end{equation}
\begin{equation}
\label{IRGsQC} 
{\cal G}^{QC}(k^{2}) = \frac{8\pi^2}{\omega^4} \, D\, {\rm e}^{-k^{2}/\omega^2} \, .
\end{equation}
Equations~(\ref{IRGs}) and ~(\ref{IRGsQC}) deliver an ultraviolet-finite model gap equation.
Hence, the regularization mass scale can be removed to infinity and the renormalization constants can be set to 1.
For the corresponding ans\"{a}tz at finite chemical potential, 
we follow that in Ref.~\cite{Chen2008}, neglecting the dependence of the effective interaction ${\cal G}$ and the gluon propagator on the chemical potential at low densities.

There are only two main parameters in our model:$D$ and $\omega$. We choose the set of values that can fit meson properties in vacuum well~\cite{Chang2009} or fit the  chiral condensate and pion decay constant $ f_{\pi}^{}$  in vacuum approximately.
We use the approximate formula for calculating $f_{\pi}^{}$ which is accurate to within $ 5\% $  in chiral limit with the rainbow approximation ~\cite{McKay1988,TA1987}.
\begin{eqnarray}
\label{fpi}
f_{\pi}^{2}&=&\int \frac{ds}{8\pi^{2}}N_{c}sB(s)^{2}[\sigma(s)^{2}_{V}\nonumber
\\ & & -2\sigma(s)_{S}\sigma(s)^{'}_{S}-2s\sigma(s)_{V}\sigma(s)^{'}_{V}\nonumber
\\ & & -s\sigma(s)_{S}\sigma(s)^{''}_{S}+s(\sigma(s)^{'}_{S})^{2}\nonumber
\\ & & -s^{2}(\sigma(s)_{V}\sigma(s)^{''}_{V}-(\sigma(s)^{'}_{V})^{2})] \, ,
\end{eqnarray}
with the primes denoting differentiation with respect to $ s=p^{2} $ , and
\begin{equation}
\label{AA}
\sigma_{V}=\frac{A(p^{2})}{p^{2}A(p^{2})+B(p^{2})} \, ,
\end{equation}
\begin{equation}
\label{BB}
\sigma_{S}=\frac{B(p^{2})}{p^{2}A(p^{2})+B(p^{2})} \, .
\end{equation}

\section{pion-nucleon sigma term}
\label{SecIII}
%With the Hellmann-Feynman theorem, pion-nucleon sigma term $\sigma_{\pi N}$ is usually given as
%\begin{equation}
%\label{sigma1}\sigma_{\pi N}=m_q\frac{\partial M_N}{\partial m_q}
%\end{equation}
%
It has been known that pion-nucleon sigma term can be determined by the chiral susceptibility
$\frac{\partial M_{N}^{}}{\partial m_{q}}$ and the current-quark-mass in the form of Eq.~(\ref{eqn:sigmaterm-definition}).
%
%is the nucleon mass. Such a definition represents the dependence of the nucleon mass on the effects of the explicit and the %dynamical chiral symmetry breaking. And it shows certainly a theoretical way to calculate the $\sigma_{\pi N}$.
%
However, it is very complicated to calculate the nucleon mass $M_{N}^{}$,
which depends on the four-dimensional Poincar\'{e}-invariant Faddeev equations in the DSE approach of QCD,
and the results are still robust to get the dependence of nucleon mass on the current-quark-mass~\cite{Ohki2008,Flambaum2005}.
Meanwhile it is oversimplified to regard the nucleon as three noninteracting constituent quarks~\cite{Fujii1995,Chang2005}.
Therefore, we do not perform the calculation from Eq.~(\ref{eqn:sigmaterm-definition}) directly.
%
%but make our evaluation based on Eq.~(\ref{cond-sigma}), which obtained from the Hellmann-Feynman theorem.
%

It has been well known that, in the QCD Hamiltonian $\hat{H}_{QCD}$, the mass term $\hat{H}_{mass}$ is
\begin{equation}
\label{hmass}\hat{H}_{mass}=\int d^3x (m_{u}^{} \bar{u}u + m_{d}\bar{d}d + \cdots  ) \, ,
\end{equation}
where $u$, $d$ denotes the up, down quark with current-quark-mass $m_{u}$, $m_{d}$, respectively,  $\cdots$ denotes the contributions from heavier quarks. It is useful to reorganize the up- and down-quark contribution to $\hat{H}_{mass}$ in order to isolate the isospin breaking effects. Defining $\bar{q} q = \frac{1}{2} (\bar{u} u + \bar{d} d )$ and $m_{q} = \frac{1}{2} ( m_{u} + m_{d})$, Eq.(\ref{hmass}) can be rewritten as
\begin{equation}
\label{hmass2}\hat{H}_{mass} \! = \! \int d^3x [2m_q\bar qq
+\frac{1}{2}(m_u-m_d)(\bar uu-\bar dd) + \cdots ] \, .
\end{equation}
Making use of the the Hellmann-Feynman theorem, one obtains (see, for example, Ref.~\cite{Cohen1992})
\begin{eqnarray}
\label{chiral}
2m_q \langle \Psi|\int d^3x \bar qq|\Psi\rangle & = & m_q
\langle \Psi \Big{|} \frac{d\hat{H}_{mass}}{dm_q} \Big{|} \Psi\rangle \nonumber\\
&=& m_q \frac{d}{dm_q}E_\Psi \, ,
\end{eqnarray}
where $|\Psi\rangle$ represents a normalized eigenvector of QCD Hamiltonian and $E_\Psi$ stands for the energy of the state $|\Psi\rangle$.

Considering the case in which $|\Psi\rangle $ is the state of hadron matter at rest with baryon number density ${n_{B}^{}}$, and also the case of the vacuum state, one has
\begin{equation}
\label{chiral2}
2 m_q [ \langle \bar{q} q \rangle_{n} - \langle
\bar{q} q \rangle_{0} ] = m_{q} \frac{d \epsilon}{d m_{q}} \, ,
\end{equation}
where $\epsilon$ is the energy density of the baryon matter and can
be written as
\begin{equation}
\label{epsilon}
\epsilon = {M_{N}^{}} {n_{B}^{}} + \delta\epsilon \, ,
\end{equation}
where $\delta\epsilon$ denotes the contributions from the kinetic energy of baryons and baryon-baryon interactions. $\delta\epsilon$ is of high order in density and is empirically small at low densities: the binding energy per nucleon at the  nuclear matter saturation density is only 16~MeV. Therefore, neglecting $\delta\epsilon$ and implementing Eq.~(\ref{chiral2}), one obtains
\begin{equation}
\label{chiral3}
2 m_{q} [ \langle \bar{q} q \rangle_{n} - \langle
\bar{q} q \rangle_{0} ] = m_{q} \frac{d M_{n}}{d m_{q}} {n_{B}^{}} =
{\sigma_{\pi N}^{}} {n_{B}^{}} \, .
\end{equation}

Replacing the baryon number density ${n_{B}^{}}$ with the quark number density ${ n_{q}^{}} = 3 {n_{B}^{}} $,
we obtain the linear dependence of the variation of the chiral condensate on the quark number density
\begin{equation}
\label{chiralvari}
[ \langle \bar{q} q \rangle_{n} - \langle \bar{q} q \rangle_{0} ] =
\frac{\sigma_{\pi N}^{}}{6 m_{q} } {n_{q}^{}}=k n_{q} \,,
\end{equation}
where $k$ is the slope which can be obtained from the linear fitting of the relation in Eq.~(\ref{chiralvari}).
Conversely, we can evaluate pion-nucleon sigma term as
\begin{equation}
\label{chiral4}  \sigma_{\pi N}^{} = 6 m_{q} k = 6 m_{q}\frac{\langle \bar{q} q\rangle_{n}-\langle \bar{q} q\rangle_{0}}
{n_q} \, .
\end{equation}
For the light u and d quarks,  we can take advantage of the Gell-Mann--Oakes--Rennerrelation~\cite{GORrelation}, which is accurate within 5\%~\cite{Jamin2002}:
\begin{equation}
\label{mq}m^{2}_{\pi}f^{2}_{\pi}=-2m^{}_{q}\langle \bar qq \rangle_{0}^{0} \, ,
\end{equation}
where $m_\pi=138 \rm{MeV}$ and $f_\pi=93\rm{MeV}$ have been well established in experiments, and $\langle \bar qq \rangle_{0}^{0}  $ is the chiral condensate in the chiral limit (represented by the superscript `0') in vacuum.
We can then obtain
\begin{equation}
\label{sigma-chlimit}
\sigma_{\pi N}^{}=3k
\Big{(}\! -\frac{m_{\pi}^{2}f_{\pi}^{2}}{ \langle \bar qq\rangle_{0}^{0}} \Big{)} \, ,
\end{equation}
with which pion-nucleon sigma term $\sigma_{\pi N}$ can be evaluated from the chiral condensate in vacuum and that in medium.

The quark number density ${n_{q}^{}}$ can be calculated from the quark propagator at finite chemical potential, $S(p;\mu)$, with the definition:
\begin{eqnarray}
\label{nq} 
{n_{q}^{}} & = &  N_{c} N_{f}Z_{2}Tr[-\gamma_{4}S(p;\mu)]\, , 
\end{eqnarray}
For light quarks, we can further approximate the  chiral condensate in Eq.(~\ref{chiralvari}) with that in the chiral limit, which
can be well defined from the quark propagator in the chiral limit:
\begin{equation}
\label{chiral00}  
-\langle \bar{q} q \rangle_{n}^{0}  =   N_{c}Z_{2} Z_{m} Tr[S^0(p;\mu)] \, ,
\end{equation}
where $Tr$ represents the trace in color and Dirac space and integration in momentum space, and $Z_{2}$, and $Z_{m}$ are renormalization constants for quark wave function and quark mass, respectively.

To be more accurate in the cases of physical u,d and even s quarks, one can take a current-quark-mass taht better fits the meson properties obtained from the BSE. However,  Eq.~(\ref{chiral00}) is divergent in the case of a finite current-quark-mass. Although some different subtraction points have been introduced to give finite values, it is still an open question to define the chiral condensate from the quark propagator with a finite quark mass: see, for example,  Refs.~\cite{Maris1997,Brodsky2010}.
Fortunately, we only need the variation of the  chiral condensate in medium, which is independent of a fixed subtraction point. Therefore, we also investigate the variation of the  chiral condensate in medium with a finite current-quark-mass, defined as
\begin{eqnarray}
\label{qu-q0}
\Delta\langle \bar{q} q \rangle_{n}^{m_{q}} & = &
\langle \bar{q} q \rangle_{n}^{m_{q}}- \langle \bar{q} q \rangle_{0}^{m_{q}}  \nonumber \\
 & = & Z_{2} Z_{m} Tr[S(p;\mu)-S(p;\mu=0)] \, , \quad
\end{eqnarray}
where the quark propagator is calculated with finite current-quark-mass $m_q$.

\begin{table*}[t]
\caption{Parameters and some characteristic numerical
results (dimensional quantities in units of MeV). DES1, DSE2, DSE3 and DSE4 are in the chiral limit $ m_q $ = 0, while DSE5 investigates the chiral condensate beyond the chiral limit, see the text for details.
\label{Table:Para1} }
\begin{center}
\begin{tabular}{|p{1.5cm}|p{1.5cm}|p{1.6cm}|p{1.5cm}|p{1.5cm}|p{1.5cm}|p{1.5cm}|p{1.5cm}|p{1.5cm}|p{1.5cm}|}\hline
~~~~DSE~&~~vertex&interaction ~&~~~~~~~$\omega$~&~~~~~~$D$~ & ~~$-\langle \bar qq \rangle^{1/3}_{0}$~ &~~~~~$ m_q $ &~~~~~~$  k $&~~~~~$  S_k $&~~~~~$ \sigma_{\pi N} $\\  \hline
~~~~DSE1~ &~~~~ RB&~~~~~GS~ &~~~~~ $500$ &~~~~$1.00$& ~~~~~$252$~&~~~~~$5.2$ &~~~~~{ $1.95$} & ~~~$0.03$
&~~~~~{ $61$ }  \\  \hline
~~~~DSE2~ &~~~~ RB&~~~~~QC  &~~~~~ $678$ &~~~~$1.10$& ~~~~~$253$~&~~~~~$5.2$ &~~~~~{ $2.04$} & ~~~$0.03$
&~~~~~{ $63$ }  \\  \hline
~~~~DSE3~ &~~~~ BC&~~~~~GS~ &~~~~~ $500$ &~~~~$0.50$& ~~~~~$258$~&~~~~~$4.7$ &~~~~~{ $2.22$} & ~~~$0.01$
&~~~~~{ $63$ }  \\  \hline
~~~~DSE4~ &~~~~ BC&~~~~~QC~ &~~~~~ $678$ &~~~~$0.55$&~~~~~$256$~ &~~~~~$4.7$ &~~~~~{ $2.21$} & ~~~$0.01$
&~~~~~{ $62$ }  \\  \hline
~~~~DSE5~ &~~~~ RB&~~~~~GS~ &~~~~~ $500$ &~~~~$1.00$&~~~~~$-$~ &~~~~~$5.2$ &~~~~~{ $1.94$}   & ~~~$0.05$
&~~~~~{ $61$ }  \\  \hline
\end{tabular}
\end{center}
\end{table*}

\section{Numerical calculations and Results}
\label{SecIV}
To carry out the numerical calculations, we need the parameters $D$ and $\omega$ in the effective interaction.
Usually the parameters are determined by fitting meson properties with the BSE approach. 
The parameters and some characteristic results at $\mu=0$ are listed in Table.~\ref{Table:Para1}.
DSE1 represents the results with the rainbow approximation and the GS model.
DSE2 represents the results with the rainbow approximation and the QC model .
DSE3 represents the results with BC vertex and the GS model.
DSE4 represents the results with BC vertex and the QC model.
DSE5 represents the results with the rainbow approximation and the GS model ,
but the variation of the chiral condensate in medium are calculated with Eq.~(\ref{qu-q0}) beyond chiral limit.

The parameters of DSE1 and DSE3 are taken from  Ref.~\cite{Chang2009}. The parameters of  DSE2 are obtained by fitting the pion decay constant $ f_{\pi}^{} = 93\,$ MeV with Eq.~(\ref{fpi}) and the  chiral condensate in vacuum
$-\langle \bar qq \rangle_{0} $.The parameter $d$ of
 DSE4 is obtained by fitting the  chiral condensate
$-\langle \bar qq \rangle_{0} $ with the same $\omega$ as in DSE2.
In DSE5, we take the same parameters as in DSE1.

\begin{figure}[htp!]
\includegraphics[width=0.46\textwidth]{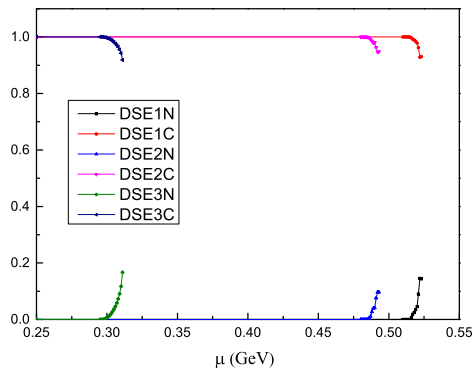}
\vspace*{-5mm}
\caption{ Quark chemical potential dependence of quark number density (DSE1N, DSE2N, DSE3N, DSE4N) (scaled with the saturation density $n_{s}^{} = 3 {n_{B,s}^{}}\approx 3 \times 0.16\,\textrm{fm}^{-3} \approx 0.0038\,\textrm{GeV}^3$) and  chiral condensate (DSE1C, DSE2C, DSE3C, DSE4C) (scaled with the values in vacuum).}
\label{chiral_cn}
\end{figure}

With the above-determined parameters and the ans\"{a}tz described in the last section, we solve the Dyson-Schwinger equation of the quark propagator and calculate the chemical potential dependence of the chiral condensate and the quark number density.
The obtained results in the chiral limit are illustrated in Fig.~\ref{chiral_cn}.\\

%\begin{figure*}
%\includegraphics[width=0.95\textwidth]{dsesigma.pdf}
%\caption{\label{fig:sigmat}(color online) Variation of the chiral %condensate in medium respect to the quark number density. }
%\end{figure*}
\begin{figure*}[htp!]
\includegraphics[width=1.0\textwidth]{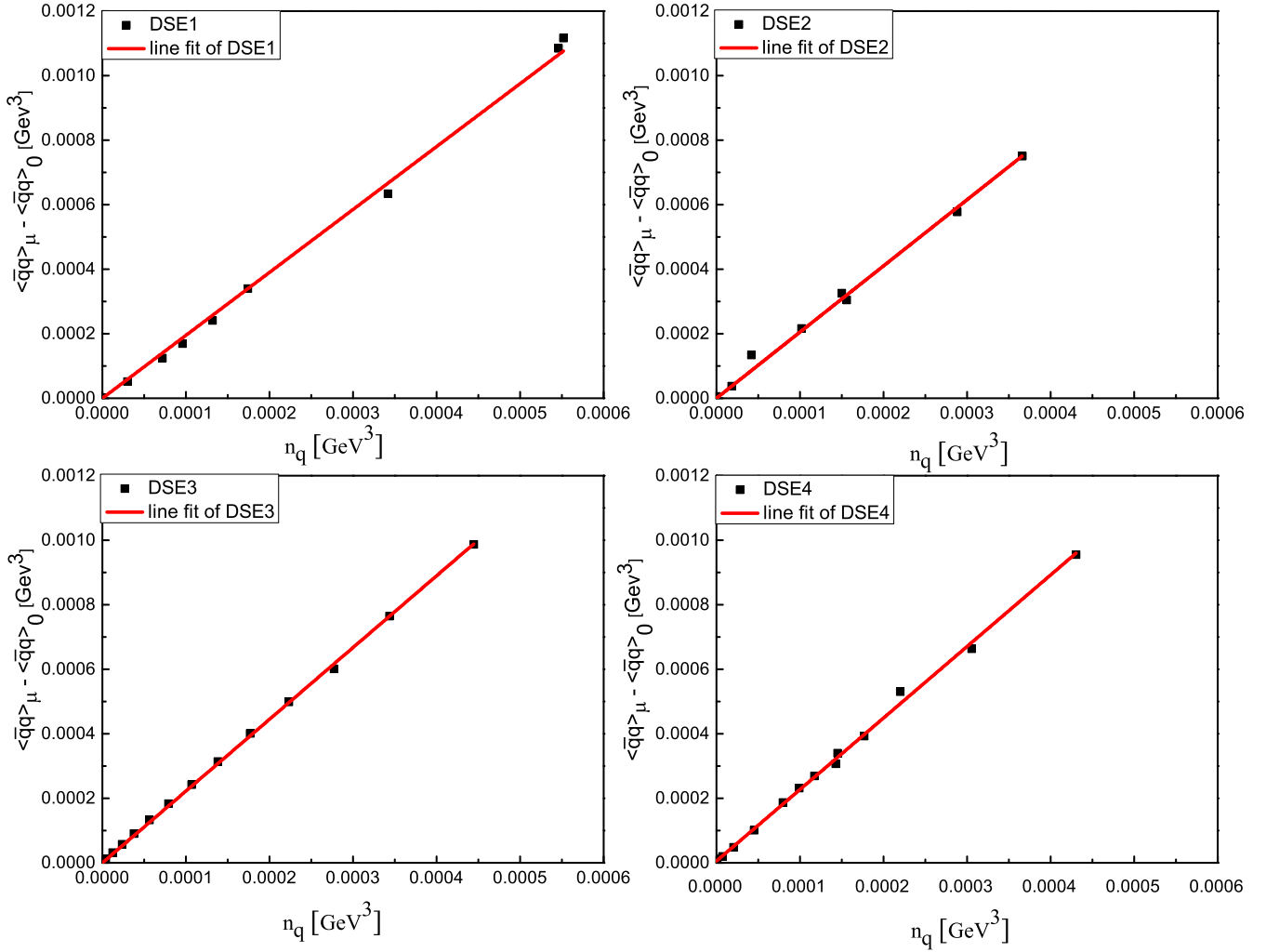}
\vspace*{-5mm}
\caption{Variation of the chiral condensate in medium respect to the quark number density.  }
\label{sigmat}
\end{figure*}

\begin{figure}[htp!]
\includegraphics[width=0.52\textwidth]{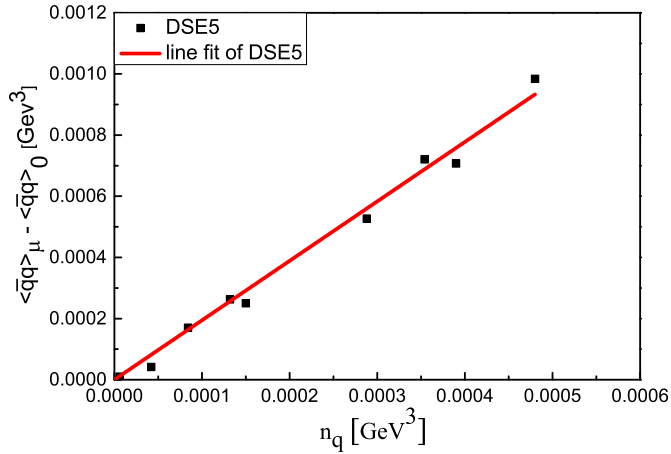}
\vspace*{-5mm}
\caption{Variation of the chair chiral condensate with resepect to the quark number density, with the model DSE5 beyond chiral limit. }
\label{dse_5s}
\end{figure}

Figuer.~\ref{chiral_cn} shows that when the quark chemical potential $\mu < M_1$, where $M_1$ is the first constituent-mass-like pole of the quark propagator, the  chiral condensate keeps the same value as that in vacuum (i.e., at $\mu=0$) and the quark number density remains zero, i.e. the system remains the same as that in the vacuum and no matter emerges\cite{Chen2008}. 
When $\mu > M_1$, the quark number density becomes nonzero and simultaneously the  chiral condensate decreases gradually. This indicates that dynamical chiral symmetry is partially restored in the medium at low density~\cite{ChangChen2007,Chen2009}.
With the above results, we get the variation of the chiral condensate  $ \Delta\langle \bar{q} q \rangle_{n}$ in the medium  with respect to the quark number density $ n_q $ of the system. The obtained result is displayed in Fig.~\ref{sigmat}.\\

Although the chemical potential dependence of quark number density and  chiral condensate is model dependent and looks complicated, Fig.~\ref{sigmat} exhibits a linear relation (in Eq.~(\ref{chiralvari})
between the variation of the chiral condensate and the baryon number density. 
We linearly fit the dependence of the variation of the chiral condensate ($y$) on quark number density ($x$) as
\begin{equation}
\label{lihe1}  
y = kx
\end{equation}
with 
\begin{equation}
\label{lih2}  
k=\frac{\sum_{i=1}^n (x_{i}-\overline{x})(y_{i}-\overline{y})}{\sum_{i=1}^n (x_{i}-\overline{x})^{2}}
\end{equation}
where $\overline{y}=n^{-1}\sum_{i=1}^n y_{i}$ is the sample average of the $y_i$ and likewise for $\overline{x}$. The standard error $S_k$ is given as:
\begin{equation}
\label{lihe4}  
S_k=\sqrt{\frac{\sum_{i=1}^n(\hat{y}_i-y_i)^2}{(n-2)\sum_{i=1}^n (x_{i}-\overline{x})^{2}}} \, ,
\end{equation}
where $\hat{y}_i=kx_i$.
 The fitted values of the slope and corresponding values of the sigma term are listed in Table.~\ref{Table:Para1}.
Remarkably, the slopes and the  
corresponding results for $\sigma_{\pi N}^{}$ quantitatively depend very weakly on the choice of the ans\"{a}tz for the quark-gluon vertex and effective interaction.

In DSE5 we investigate the effect of a finite current-quark-mass with Eq.~(\ref{qu-q0}),
and one can find from Fig.\ref{dse_5s} and Table. \ref{Table:Para1} that the results of $ k $ and the sigma term are not sensitive to such a change.
Therefore, we can be quite confident on the chiral limit approximation ~(\ref{sigma-chlimit}) for the pion-nucleon sigma term and the validation of Eq.~(\ref{qu-q0}) in the case of a finite current-quark-mass.\\

\begin{figure}[htp!]
\includegraphics[width=0.63\textwidth]{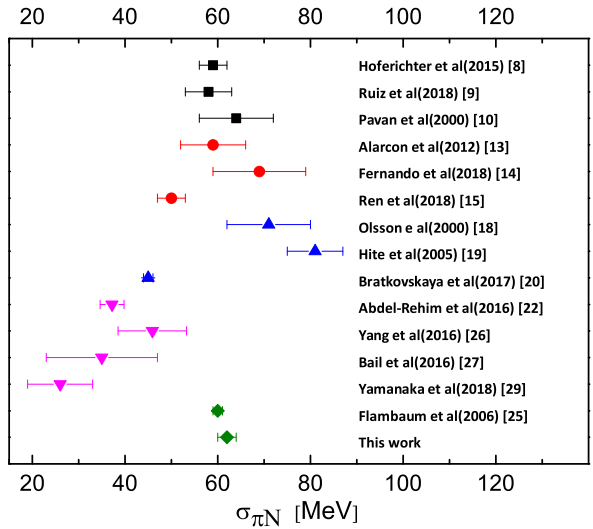}
\vspace*{-5mm}
\caption{Comparison of our result for pion-nucleon sigma term $\sigma_{\pi N}^{}  $ with those in the literature, from $ \pi$-$N $ scattering data (black)\cite{Pavan1999,Hoferichter2015,Ruiz2018}, chiral perturbation theory (red)\cite{Camalich2012,Fernando2018,Ren2018}, QCD (magenta )\cite{Abdel2016,Yang2016,Bail2016,Yamanaka2018}, from various other models (blue)\cite{Olsson2000,Hite2005,Bratkovskaya2017} and the DSE(green) approach\cite{Flambaum2005}.}
\label{result}
\end{figure}

With the above results, we estimate that pion-nucleon sigma term $\sigma_{\pi N}^{} $ is about $6 k \approx 12 $ times the current-quark-mass $m_q$.
This leads to a larger value of $\sigma_{\pi N}^{}  $ than the result given in Ref.~\cite{Chang2005},
which estimated that $\sigma_{\pi N}^{}  $ in the  chiral limit in the vacuum is 9/2 times $m_q$.
Finally, 
we obtain pion-nucleon sigma term $\sigma_{\pi N}^{}  $ = 62(1) MeV,
where the (1) represents the systematic error due to our different ans\"{a}tz for the quark-gluon vertex and gluon propagator.
Comparing our results with recent experimental data and theoretical results in Fig.~\ref{result},
one can note that our present result is remarkably consistent with the recent experimental results.
%
%With such a larger value of $\sigma_{\pi N}$, the  chiral condensate in medium decreases faster than those given in Refs.~\cite{Cohen1992，Tsushima2007}. From a direct extrapolation, the  chiral condensate in medium at the saturation density of nuclear matter decreases to about half of that in vacuum.

\section{Summary and Remark}
\label{SecV}
In summary, using the Dyson-Schwinger equation approach of QCD, we calculated the  chiral condensate in strong-interaction matter at low density, and then evaluated pion-nucleon sigma term $\sigma_{{\pi} N}^{}$ via the Hellmann-Feynman theorem.
In this work, we adopted various ans\"{a}tz for the gluon propagator and quark-gluon vertex,
and found that our evaluated value for the pion nucleon sigma term depends on the model very weakly.
We obtained the result $\sigma_{\pi N}^{}=62(1)\,$ MeV in the Dyson-Schwinger equation approach of QCD.
Our results are rather consistent with the relatively large value given in recent experimental analyses.

In solving the Dyson-Schwinger equation of the quark propagator,
we adopted models of the effective interaction (gluon propagator) that are independent of the quark chemical potential.
However, the gluon propagator should depend on the quark chemical potential via the quark-loop diagram in its vacuum polarization. It is reasonably to expect this to be a lower-order effect on pion-nucleon sigma term.
However, the situation may be different when calculating the sigma term for the strange quark $\sigma_s$,
which is important for dark matter searches~\cite{Giedt2009}.
Since the strange quark chemical potential is zero at low baryon number densities,
such an effect is the leading-order effect for the variation of the strange chiral condensate.
It is then necessary to consider this effect when calculating $\sigma_{s}$.
We could investigate it to calculate  $\sigma_s$ and further improve our results for $\sigma_{\pi N}^{}$.
On the other hand, the more sophisticated quark-gluon vertex has also been established (e.g., Refs.~\cite{CLR,QCLR,TGL}).
Calculating $\sigma_{\pi N}$ and $\sigma_{s}$ with the more sophisticated quark-gluon vertex is also interesting and related work is in progress.

%\bigskip
\begin{acknowledgments}

%
%10275002, 10435080, the Major State Basic Research Development
%Program under contract No. G2000077400, the research foundation of
%the Ministry of Education, China (MOEC), under contact No. 305001
%and the Doctoral Program Foundation of MOEC under grant No.
%20040001010. One of the authors (YXL) thanks also the support of the
%Foundation for University Key Teacher by the MOEC.
%
 We acknowledge financial support from National Natural Science Foundation of China (Grants No. 11475149, and No. U1738130).
We are also in debt to Prof. Yu-xin Liu at Peking University for his stimulating discussions. 
\end{acknowledgments}

%\bibliography{reference}

\bibliography{reference}

\end{document}